# Automatic target validation based on neuroscientific literature mining for tractography

Xavier Vasques[1,2,3†], Renaud Richardet[1†], Sean L. Hill[1], David Slater[4,5], Jean-Cedric Chappelier[6], Etienne Pralong[5], Jocelyne Bloch[5], Bogdan Draganski[4,5] and Laura Cif[4,5,7]*

[1] Blue Brain Project, Brain Mind Institute, Ecole Polytechnique Fédérale de Lausanne, Lausanne, Switzerland, [2] IBM Systems, France, [3] Laboratoire de Recherche en Neurosciences Cliniques, France, [4] Laboratoire de Recherche Neuroimagerie, Université de Lausanne, Lausanne, Switzerland, [5] Département des Neurosciences Cliniques, Centre Hospitalier Universitaire Vaudois, Université de Lausanne, Lausanne, Switzerland, [6] School of Computer and Communication Sciences, Ecole Polytechnique Fédérale de Lausanne, Lausanne, Switzerland, [7] Département de Neurochirurgie, Hôpital Gui de Chauliac, Centre Hospitalier Régional Universitaire de Montpellier, Université Montpellier 1, Montpellier, France

Target identification for tractography studies requires solid anatomical knowledge validated by an extensive literature review across species for each seed structure to be studied. Manual literature review to identify targets for a given seed region is tedious and potentially subjective. Therefore, complementary approaches would be useful. We propose to use text-mining models to automatically suggest potential targets from the neuroscientific literature, full-text articles and abstracts, so that they can be used for anatomical connection studies and more specifically for tractography. We applied text-mining models to three structures: two well-studied structures, since validated deep brain stimulation targets, the internal globus pallidus and the subthalamic nucleus and, the nucleus accumbens, an exploratory target for treating psychiatric disorders. We performed a systematic review of the literature to document the projections of the three selected structures and compared it with the targets proposed by text-mining models, both in rat and primate (including human). We ran probabilistic tractography on the nucleus accumbens and compared the output with the results of the text-mining models and literature review. Overall, text-mining the literature could find three times as many targets as two man-weeks of curation could. The overall efficiency of the text-mining against literature review in our study was 98% recall (at 36% precision), meaning that over all the targets for the three selected seeds, only one target has been missed by text-mining. We demonstrate that connectivity for a structure of interest can be extracted from a very large amount of publications and abstracts. We believe this tool will be useful in helping the neuroscience community to facilitate connectivity studies of particular brain regions. The text mining tools used for the study are part of the HBP Neuroinformatics Platform, publicly available at http://connectivity-brainer.rhcloud.com/.

# Introduction

Determining the wiring diagram of the human brain is one of the greatest challenges in neurosciences (Sporns, 2011). In initiatives such as the Human Connectome Project (HCP) (www.humanconnectome.org), tractography occupies a key place in establishing the structural basis of the human connectome. Diffusion tensor imaging (DTI) has been introduced to document and measure *in vivo* anatomical connectivity between regions (Jbabdi and Johansen-Berg, 2011). DTI offers an overall view of brain anatomy, including the pattern and degree of connectivity between different regions, raising immediate hypothesis for brain function and for clinical applications such as deep brain stimulation (DBS) (Coenen et al., 2011, 2012a,b). DBS is a therapeutical approach for movement (Pouratian et al., 2011; Rozanski et al., 2014; Sweet et al., 2014a,b) and psychiatric disorders (Lujan et al., 2008; Lakhan and Callaway, 2010; Lehman et al., 2011), targeting different basal ganglia structures and delivering chronic stimulation to them (Barkhoudarian et al., 2010; Sedrak et al., 2010; Traynor et al., 2010; Taljan et al., 2011; Lambert et al., 2012; Chowdhury et al., 2013). In combination with other technologies, DTI represents a powerful tool providing further insight on the networks influenced by neuromodulation (Barkhoudarian et al., 2010; Chaturvedi et al., 2010; McIntyre and Foutz, 2013; Howell et al., 2014) and consequently a better understanding of the mechanism of action and effects of DBS.

One of the major limitations of tractography is related to its outputs because of, potential underestimates of the fiber tracts when compared to other methods (Ciccarelli et al., 2003a,b; Kinoshita et al., 2005) such as fiber pathways that are reported in dissection and tracer studies that are absent in diffusion tensor tractography studies (Behrens et al., 2007). Therefore, responsible use of tractography requires careful consideration of the scope and limitations of the different techniques (Johansen-Berg and Behrens, 2006), knowing that observations are only fraction of the reality. Probabilistic tractography approach, as opposed to deterministic approach, depicts more fibers, thus leading to a more limited underestimation, since it assumes a distribution of orientation, as opposed to a single orientation at each voxel. Local tractography fits pathways step by step and is suitable for exploratory studies of connections compared with global tractography, and is more suitable for reconstruction of known white matter pathways. It is essential to have a thorough previous knowledge of the connections between the regions under investigation in order to validate the relevant fibers depicted via tractography, to pinpoint misses and for the choice of the method to be used. Mainly two approaches are used in probabilistic tracking (Catani et al., 2002; Wakana et al., 2007). In the first approach, all fiber tracts are obtained through a single seed region of interest (ROI) such that only fibers passing through the seed are included in the reconstructed tract. In the second one, the knowledge-based multiple-ROI approach, all fiber tracts are obtained through a seed to target ROIs, with logical and concatenation of two ROIs, such that only fibers passing through both ROIs are included in the reconstructed tract. Obviously spurious fibers are removed from the fiber tract by using an additional avoidance ROI (logical NOT operation) (Wakana et al., 2007). In the first approach, we only have to create a mask of the ROI (automatically or manually), in order to generate a connectivity distribution from the specified region of interest. Probabilistic tractography is performed from every voxel with a value greater than 0 in this mask. The output file is a single image in the space of the specified seed mask. All brain voxels have a value (though many of these may be zero) representing the number of samples that pass through that voxel from the seed mask. Target identification is a further crucial step for guided tractography from a seed region, to estimate the probability of their interconnection. Target identification requires solid anatomical knowledge documented by an extensive literature review across species for each seed structure to be studied. Existing literature in human is often conflicting and limited. Furthermore, experiments studying connectivity between individual brain regions are not reported in a normalized, structured and centralized repository, but published in plain text, scattered among individual scientific publications (Richardet et al., 2015). Consequently, manual literature review (LIT) to identify targets for a given seed region is tedious and potentially subjective. Therefore, complementary approaches would be very useful for the neuroscience community.

In this article, we propose to use text-mining (TM) models to automatically generate potential targets from the neuroscientific literature, so that they can be used for anatomical connection studies and more specifically for tractography studies. These TM models aggregate brain region connectivity from a very large amount of published neuroscience full-text articles and PubMed abstracts. To illustrate and evaluate the methodology, we applied TM models to three structures: two well-studied structures, since validated DBS targets for movement disorders, the internal globus pallidus (GPi) and the subthalamic nucleus (STN) and, the nucleus accumbens (NAcc), exploratory target for treating psychiatric disorders. We performed a systematic review of the literature to document the projections of the three selected structures and compared it with the structures proposed by TM models, both in rat and primate (including human). To assess the results of the TM models, a comparison has been made between the two methods for the well-described GPi and STN. Finally, we ran probabilistic tractography on the NAcc and compared the output with the results of the TM models and literature review. The objective of this paper is to document/support the validity of the TM models approach in helping to identify the targets to be explored for a given seed structure in (probabilistic) tractography projects.

# Materials and Methods

## Search Strategy for Identification of the Three Seed Structures and Their Connections in Rat and Primates, Including Humans

Relevant publications were obtained using the PubMed database and references from the consulted articles. The PubMed database was manually searched for articles describing connections of the three nuclei, globus pallidus internus, subthalamic nucleus, and nucleus accumbens. MeSH headings used were "globus pallidus," "entopeduncular nucleus" (corresponding to the medial segment of the globus pallidus in rats), "subthalamic nucleus," and "nucleus accumbens." We further searched for the following terms: "globus pallidus internus," "pallidum internum," "internal globus pallidus," "globus pallidus pars interna," and "medial globus pallidus." We combined them with the following MeSH headings for the studied species: "rats," "primates," and "human" and with the following key words: "connections," "projections," "afferents," and "efferents." Only articles written in English were reviewed. We used Terminologia Anatomica as reference for official nomenclature of the studied regions and structures.

## Automatic Information Extraction from the Neuroscientific Literature

To accelerate manual literature search, we used TM methods that distill very large amount of scientific articles in order to extract brain regions that are potentially connected. The TM process consist of three phases: first, identifying mentions of brain regions in text; second, determining which of these brain regions are connected, and third, aggregating and reporting on potential connections in a database easily searchable by neuroscientists. For the identification of brain regions, two complementary named entity recognizers (NER) were developed. The first NER uses a lexicon of all 1197 brain regions from the Allen Mouse Brain Atlas (ABA) (http://www.brain-map.org) that is automatically augmented with corresponding synonyms found in several lexica (Richardet et al., 2015) of rodent brain region: the Brain Architecture Management System (BAMS) (Bota and Swanson, 2008), Neuronames (Bowden and Martin, 1995; Bowden and Dubach, 2003), Paxinos and Watson (Paxinos and Watson, 2007), Swanson (Puelles Lopez, 2000).

The second NER (BrainNER) relies on a machine-learning model (linear chain conditional random field) trained on WhiteText, a manually annotated corpus of 18,242 brain region mentions (French, 2009; French et al., 2012). The advantage of this statistical approach is that the model will match complex brain region names, even if they are not present in a lexicon, for example "contralateral prepositus hypoglossal nucleus" or "distal parts of the inferior anterior cerebellar cortex."

Once brain regions were identified, the second step was to determine whether two brain regions mentioned in a sentence were anatomically connected or not. To this end, three different models were combined: (1) FILTER considers all possible brain region co-occurrences, and subsequently applies filters to remove unlikely ones; (2) KERNEL relies on a supervised machine-learning classifier; (3) RULES consist of 9 manually crafted rules of the kind "*projection from the region A to the region C and the region D.*" The resulting database shows, by selecting a region of interest, all other connected regions extracted from the literature and the possibility to drill down to the individual sentences for detailed analysis. The complete methodology can be found in Richardet et al. (2015).

The database is publicly accessible through a simple and intuitive web application. This application provides a matrix of brain regions co-occurences displaying the top N regions for which the most connection mentions was found (see **Supplementary Figure 1**). All matrix values are linked to the corresponding detailed list of sentences from neuroscientific articles. For example, **Supplementary Figure 3** displays the extracted sentences between the Allen Brain Atlas regions "Periaqueductal gray" and "Nucleus accumbens." Each sentence is itself linked to PubMed so that the user can go back to the original article. Additionally, the user has the ability to provide feedback by either validating the sentence or rejecting it. Finally, it is possible to search for one particular brain regions of interest, and then list all the other brain regions potentially connected to it (for which connectivity events have been found in the literature), see **Supplementary Figure 2**. The web application also exposes a REST API to interact with the extracted connectivity programmatically.

## Guided Probabilistic Tractography of Nucleus Accumbens

High-resolution multi-parameter quantitative MRI (MPM) and high angular resolution diffusion imaging (HARDI) were acquired on a 3T whole-body MRI system (Magnetom Prisma, Siemens Medical System, Germany). The quantitative MPM acquisitions consisted of three multi-echo 3D fast low angle shot (FLASH) with proton density (PD), magnetization transfer (MT) and T1 weighted contrast as described elsewhere (Helms et al., 2008), whole brain coverage, 1 mm$^3$ resolution, FOV: (240, 176, 256) mm along A-P, L-R, H-F directions. Since previous research demonstrated that MT saturation maps provide better contrast in subcortical structures compared to T1w images (Helms et al., 2008), MT saturation maps were used for delineation of the NAcc. For the diffusion weighted acquisition we used a HARDI protocol with 60 gradient directions at $b$-value = 2000 s.mm$^{-2}$ and 13 interleaved b0 images. The following acquisition parameters were set: TE/TR = 69/7400 ms; 2 × 2 × 2 mm isotropic resolution with 70 axial slices; FoV read = 192 mm, FoV phase = 212 mm; matrix size 96 × 106; and GRAPPA factor 2. The study collecting imaging data in healthy subjects and disease conditions was approved by the Commission cantonale (VD) d'éthique de la recherche sur l'être humain, Switzerland (Protocole 207/10). Informed consent was obtained from all subjects.

HARDI preprocessing included motion correction, eddy current and correction of the vectors using the Artifact correction in diffusion MRI (ACID) toolbox (Mohammadi et al., 2010) into the batch system of Statistical Parametric Mapping 8 (SPM8). We used FSL and FMRIB's Diffusion Toolbox (FDT) (www.fmrib.ox.ac.uk/fsl) to perform segregation of brain tissue from non-brain tissue using the Brain Extraction Tool, local fitting of diffusion tensors and construction of individual FA maps using DTIFIT, and tensor estimation with BEDPOSTX routine with the following options: Fibers (3), Weight (1), Burn In (1000). A segregation of brain tissue from non-brain tissue using the Brain Extraction Tool (BET) was performed on the structural images. The left and right NAcc were segmented using FIRST from FSL on structural images. After the segmentation, we applied boundaries correction using first_boundary_corr that is used for the classification of the boundary voxels in the volumetric output for a single structure. We applied *fast* boundary correction method that used FSL's FAST-based tissue classification for correcting boundary voxels. The results were visualized and checked using Freesurfer (Freeview) image analysis suite (Version 5.1.0) (http://surfer.nmr.mgh.harvard.edu/). The same method has been applied to extract sub-cortical target masks. The other target masks were extracted with Individual Brain Atlases using SPM (IBMASPM) (Aleman-Gomez et al., 2006). The masks have been extracted using ITK-SNAP (http://www.itksnap.org). All the masks have been binarized. Masks have been taken by manually drawing the region of interest, when automatic extraction results did not reach quality expectations or when region masks were not available within the previously mentioned tools. This was the case for subthalamic nucleus, substantia nigra, ventral tegmental area, hypothalamus, habenula and subcalosal cingulate (Chowdhury et al., 2013). Registration from structural to diffusion space was performed using FMRIB's Linear Image Registration Tool (FLIRT) in FSL. We performed a 3D-to-3D registration between diffusion and structural image with the affine model, 12° of Freedom and the use of the Tri-linear final interpolation method. The resulting structural to diffusion registration was manually checked to ensure satisfactory alignment, with particular attention paid to the regional borders of the BG in general and NAcc, in particular. We run tractography analysis with probabilistic tracking (probtrackx) in FSL using the segmented left and right NAcc seeds, the target masks and the output matrix from the registration (structural to diffusion). We used the following parameters: curvature threshold of 0.2 corresponding to a minimum angle of approximately 80°, number of samples 5000 and, loopcheck option. We developed a Matlab script in order to extract from the probabilistic tractography outputs the number of tracts that leave a voxel from NAcc to reach a given target. A Python script has been developed to calculate the number of voxels within the NAcc that have a probability greater than 1% to be connected to a specific target. Voxels with a connection probability of at least 0.01 were included as voxels containing anatomically valid pathway. The 1% threshold is a typical threshold used in probabilistic tractography (Lambert et al., 2012; Li et al., 2013). Threshold was set to 1% (out of the 5000 generated from each seed voxel) to reject voxels with low probability. This means that at least 1% of the identified fibers intersect the voxels in the valid pathway. This gave us a matrix of "strengths" of interconnection on a scale of 0–100%.

We built up the NAcc connectivity maps, with the associative map corresponding to the NAcc putative core [the prefrontal cortex including frontal superior, frontal middle and frontal inferior opercular and triangular gyri as well as the lateral orbitofrontal (l-OFC) cortex] and the limbic map corresponding, to the NAcc putative shell (distribution to medial OFC, frontal inferior orbital, frontal superior orbital, anterior cingulate cortex, subcalosal cingulate cortex, amygdala, hippocampus, habenula, hypothalamus, and ventral tegmental area) (Kopell and Greenberg, 2008; Baliki et al., 2013).

# Results

## Manual Literature Review (LIT)

The literature review has been performed by two of the authors (LC and JB) and took approximately 5 working days for the three regions. Below follows a detailed description of the three seed structures and their connections in rat and primates, including humans, based on a systematic review of the literature. The summary of the systematic review is presented in **Table 1**.

## Internal Globus Pallidus

The globus pallidus is composed by two segments, a lateral, larger segment, the external globus pallidus (GPe) and a smaller, medial segment, the GPi. Furthermore, according to its vertical orientation, the subcomissural part of the globus pallidus is known as the ventral pallidum. In rat, the internal segment is called the entopeduncularis nucleus, the globus pallidus referring only to the external globus pallidus. In human, the two segments are separated by the medial medullary lamina. The GPi is further subdivided into a medial (GPi-m) and a lateral segment (GPi-l) by the accessory medullary lamina. Since the GPi is the DBS target for treating movement disorders, we will further focus only on the GPi.

TABLE 1 | Summary of the manual literature review.

| Afferents | Efferents |
|---|---|
| **GLOBUS PALLIDUS INTERNUS** | |
| Subthalamic nucleus | Thalamus |
| Substantia nigra pars compacta | Lateral habenula |
| Ventral tegmental area | Substantia nigra |
| Neostriatum | Pedunculopontine nucleus |
| | Cerebral cortex (rat) |
| | Neostriatum |
| **SUBTHALAMIC NUCLEUS** | |
| Primary motor cortex | Globus Pallidus internus |
| Supplementary motor area | Globus Pallidus externus |
| Frontal eye field | Substantia nigra pars compacta |
| Somatosensory cortex | Substantia nigra pars reticulata |
| Anterior cingulate | Ventral thalamic nuclei ipsilaterally |
| Globus Pallidus externus | Parafascicularis thalamic nucleus contralaterally (rat) |
| Substantia nigra pars compacta | Substantia innominata |
| Ventral tegmental area | Ventral pallidum |
| Dorsal raphe nucleus | Pedunculopontine nucleus |
| Pedunculopontine nucleus | Ipsilateral cortex (rat) |
| Centro-median/parafascicularis complex | Neostriatum (rat) |
| | Spinal cord (rat) |
| **NUCLEUS ACCUMBENS** | |
| Orbitofrontal cortex | Ventral pallidum |
| Anterior cingulate | Substantia nigra pars compacta |
| Subgenual cortex | Substantia nigra pars reticulate |
| Pregenual cortex | Ventral tegmental area |
| Hippocampus | Hippocampus |
| Parahippocampal cortex | Caudate |
| Amygdala | Putamen |
| Substantia nigra pars compacta | Medio-dorsal thalamus |
| Ventral tegmental area | Cingulate gyrus |
| | Substantia innominata (rat) |
| | Lateral preoptic area (rat) |
| | Lateral hypothalamic area (rat) |

In rat, the two major afferents of the entopeduncular nucleus are the neostriatum and the STN, which have opposing physiological effects on entopeduncular neurons. The striato-fugal fibers project to the entopeduncular nucleus as well as to substantia nigra, although the majority of the fibers terminate in the globus pallidus (Wu et al., 2000). Topographical and synaptic organization of the so-called direct (neostriatum to entopeduncular nucleus) and indirect pathways (involving the STN and the globus pallidus) is capable of mediating the inhibition and excitation of output neurons in the entopeduncular nucleus (Bevan and Bolam, 1995). Reciprocal connections between internal and external segments have been identified (Kincaid et al., 1991a,b) and between the pallidal complex and the STN (Smith and Bolam, 1991). A projection from the NAcc to the entopeduncular nucleus terminates in its antero-ventral (subcomissural) part (Mogenson and Nielsen, 1983; Mogenson et al., 1983). The ventral pallidum receives substantial input from the ventral tegmental area (VTA) (Napier and Maslowski-Cobuzzi, 1994). Other afferent projections to the globus pallidus and entopeduncular nucleus as well as to the ventral pallidum have been described, from the cortex (Naito and Kita, 1994), thalamus (parafascicular nucleus), dorsal raphe nucleus (Kincaid et al., 1991a,b). The entopeduncular nucleus projects mainly to the thalamus, the ventrolateral (VL), ventromedial (VM), medial dorsal, and centromedian-parafascicular complex, but also to the lateral habenula, the pedunculopontine nucleus, and the frontal cortex (Kha et al., 2000).

In non-human primate and human, afferents to the GPi are constituted by the projections of the striatal medium spiny neurons (representing the direct pathway) (Haber et al., 1990a,b) that will converge toward the GPi and by the neurons of the subthalamic nucleus. Both, caudate and putamen project to the GPi. The ventral striatum that includes NAcc projects to the ventral or limbic pallidum, including the rostral to the anterior capsule region of the globus pallidus. The projections from the neostriatum including NAcc use gamma-amminobutyric acid (GABA) as neurotransmitter and are supposed to be inhibitory. The subthalamo-pallidal projection is excitatory and glutaminergic (Smith and Parent, 1988). Dopaminergic projections from the substantia nigra (SN) and ventral tegmental area (VTA) have been demonstrated and these fibers pass to both, GPi and GPe. The major output arising exclusively from the GPi is to the thalamus (Hazrati and Parent, 1991) and the pedunculopontine nucleus (PPN) (Parent and Cicchetti, 1998). It has been suggested that the GPi has two distinct sites of origin of efferent fibers: a central "motor" zone sending axons to the thalamus, mainly the ventro-lateralis anterior nucleus following nomenclature of Jones (Jones, 1990), to the supplementary motor cortices and, the PPN. The second zone, the "peripheral" limbic zone, projects to the lateral hypothalamus and habenula (Parent, 1979), the STN and SN (Parent and De Bellefeuille, 1983; Parent et al., 1984) and to the prefrontal cortex via the dorsomedial nucleus of thalamus. In humans, when functional neurosurgery is proposed for movement disorders, the sensorimotor GPi is targeted at the posteroventral and lateral aspect of the nucleus (Laitinen et al., 1992; Coubes et al., 2004). The centro-median/parafascicular (CM/Pf) complex receives a substantial innervation from the GPi (Baron et al., 2001; Sidibe et al., 2002). Pallidal neurones project to a lesser degree, to the nucleus ventralis anterior. The associative and limbic areas of the GPi also project to the PPN (Shink and Smith, 1995). The majority of this information is derived from primate studies. The pallido-thalamic projection is mainly inhibitory and GABA-ergic.

## Subthalamic Nucleus

The STN is located within the caudal part of the diencephalon, between the ventral part of the zona incerta and the dorsal portion of the cerebral peduncles, ventral to the thalamus and lateral to the hypothalamus, parallel to the internal capsule, placed medially to the apex of globus pallidus. STN receives direct glutaminergic cortical projections as well as from the intralaminar thalamic nuclei (mainly ipsilateral but also contralateral). The main afferents to the STN are the cortico-subthalamic projections and the pallido-subthalamic pathways.

In rat, the STN receives massive cortical projections from the primary motor, prefrontal, anterior cingulate, primary somatosensory cortices (Kitai and Deniau, 1981). Pallido-subthalamic fibers arise from the globus pallidus (Smith and Bolam, 1990a,b; Kita and Kitai, 1994). The nigro-subthalamic pathway arises from SN and retrorubal and ventral tegmental areas (Hassani et al., 1997), providing dopaminergic innervation. Thalamo-subthalamic projections arise from the CM/Pf complex passing through zona incerta to reach the ipsilateral rostral STN. This pathway is demonstrated in rat (Sugimoto and Hattori, 1983; Sugimoto et al., 1983) but its role in humans remains uncertain. Other projections originate within the dorsal raphe nucleus and PPN (Canteras et al., 1990; Bevan et al., 1994a,b, 1995a,b; Bevan and Bolam, 1995). STN efferent projections are directed toward the basal ganglia nuclei. In rat, STN efferents are directed toward the GP and the SN pars reticulata but also pars compacta. Furthermore, STN projects to the thalamic ventral motor nuclei ipsilaterally and to the parafascicularis nucleus contralaterally. Further projections of the STN have been described to substantia innominata, ventral pallidum, PPN, neostriatum, ipsilateral cerebral cortex (Degos et al., 2008) and the spinal cord.

In non-human primate and human, a monosynaptic cortical connection has been described as the hyperdirect pathway originating within the primary motor cortex, the supplementary motor area and the frontal eye field and conveying the information from cortex to the GPi more rapidly than via the cortico-striato-pallidal route (Nambu et al., 2000). The GPe projects to the subthalamic neurons using GABAergic transmission. This projection is supposed to be inhibitory and belongs to the indirect pathway. The nigro-subthalamic pathway arises from SN pars compacta (Lanciego et al., 2012) retrorubal area and VTA providing dopaminergic innervation which in humans may be by the way of the dopamine D1 receptors (Augood et al., 2000). Most STN efferent neurons send axons that simultaneously innervate the GPi, GPe, and SN pars reticulata (Nauta and Cole, 1978; Rico et al., 2010). In addition to STN projections to the GPi, GPe, and SN pars reticulata, efferent STN neurons also innervate thalamic targets, ipsilateral ventral thalamic motor nuclei (Nauta and Cole, 1978; Rico et al., 2010) and contralateral parafascicular nucleus. Furthermore, dual retrograde tract-tracing studies have shown that subthalamic projections reaching the GPi and ventral thalamic nuclei arise from different subpopulations of STN neurons (Rico et al., 2010).

### Nucleus Accumbens

NAcc together with the ventral part of the caudate and of the putamen constitute the ventral striatum. The anatomical continuity between NAcc and the structures of the extended amygdala, the ventral pallidum and nucleus basalis of Meynert illustrate the strong relationship between the ventral subcomissural part of the basal ganglia (BG) and the subcortical limbic system, rendering precise delimitation of them challenging. A topographic subdivision of the NAcc into shell and core region has been described (Voorn et al., 1989; Heimer et al., 1997; Zahm, 1999), sharply marked in rodents (Meredith et al., 1996) but more challenging to identify and delineate in primates and human, in whom several different histochemical markers must be associated (Meredith et al., 1996; Brauer et al., 2000). The shell represents the ventral and medial part and the core the dorsal and central part of the nucleus. Nevertheless, significant differences exist between location and connections in rat and primates and more specifically in human. As for the striosome/matrix subdivision for the striatum, the core/shell subdivision is relevant for the information processing within the BG since each of the compartments have at least partially distinct cortical afferents. Overall, afferents to NAcc originate in the hippocampus, and prefrontal areas such as the orbitofrontal cortex and anterior cingulate. Other projections originate in subcortical structures, including amygdala.

In rat, the core receives projections from the dorsal part of the medial prefrontal cortex (corresponding to the dorsal prelimbic and anterior cingulate cortex) and from the parahippocampal cortex, while the shell receives projections mostly from the ventral parts of the medial prefrontal cortex (corresponding to the infralimbic and ventral prelimbic cortices) (Berendse et al., 1992a,b). The major part of the amygdalar projections to the BG is to the NAcc (McDonald, 1991), different for the core/shell subdivisions, respectively. The core receives projections from the anterior part of the basolateral amygdala via the ventral amygdalo-fugal pathway while the shell receives afferences from its posterior aspect and from the central nucleus of amygdala via the sublenticular and supracapsular parts of the extended amygdala (Alheid et al., 1998). The shell also receives afferences from the hippocampus (Kelley and Domesick, 1982). The core projects mainly to the dorsal subcomisural part of the ventral pallidum. The shell project to the ventral and medial part of the ventral pallidum, to the hypothalamus and the the mesencephalic dopaminergic neurons (VTA and SNc) (Berendse et al., 1992a; Heimer et al., 1997). Fibers from NAcc also pass to subpallidal structures including the substantia innominata (Berendse et al., 1992a), lateral preoptic and lateral hypothalamic area (Mogenson et al., 1983).

In primate and human, the literature reporting on NAcc connections is poorer and subjective, since it is based mainly on data from rodents and non-primate mammalians. The equivalent of the shell would receive predominant afferences from the subgenual cortices in comparison to the orbitofrontal cortex, while the core would receive similar projections from these different regions (Haber et al., 2000). NAcc, especially the putative shell region, receives a strong dopaminergic input from the VTA and from the dorsal tier of the substantia nigra (mainly the putative core) (Haber et al., 2000; Haber, 2003). Based on rodent studies, one can hypothesize that NAcc afferents are provided by the baso-lateral amygdala and most probably also the central and medial amygdalar nuclei. NAcc main efferents innervate the pallidum, striatum, mediodorsal thalamus, prefrontal, including cingulate cortex and the mesolimbic dopaminergic areas (Baliki et al., 2013). The putative core projects mainly to the dorsal subcomisural part of the ventral pallidum. The core also projects to the ventromedial SN pars compacta but also to more lateral aspects of the substantia nigra. The shell would project to the ventral and medial part of the ventral pallidum, to the hypothalamus and the VTA, as well as to the SN pars reticulata.

### Text-Mining (TM)

TM models were evaluated at different levels. First, the two NERs and three extractors are evaluated against a manually annotated corpus. Second, the complete system is evaluated against *in-vivo* connectivity from ABA. The TM models were then applied on two large corpora, and the extracted brain regions and connections are discussed. Last, we compared and analyzed the results between TM and LIT for the three structures.

The precision of both NERs was estimated on the WhiteText annotation corpus and is 84.6% (BraiNER), meaning that 85 out of 100 brain regions are correctly identified. The performance of all three extractors was evaluated on 3097 manually annotated connectivity relations, reaching a precision of 45, 60, and 72%, respectively. The resulting database contains over 4 million (lexical) and 4.5 million (machine learning) brain region mentions, and over 100,000 (lexical) and 460,000 (machine-learning) potential brain region connections. The complete system was evaluated against *in vivo* connectivity data from ABA with an estimated precision of 78% for the brain region connections that were found in the literature (recall could not be evaluated). This means that almost 8 out of 10 connections predicted by the TM system have also been experimentally measured *in vivo*. **Table 2** provides the statistics of the corpora used, extracted brain regions and connections (Richardet et al., 2015).

**Table 3** lists potential targets for the GPi and STN, as provided by the TM models. The potential targets are ranked by their decreasing score, the score representing the rounded number of connection mentions, normalized by the confidence[1] that each connection has been extracted correctly. Therefore, a high score means that many articles have been found. We stress the fact that the frequency of a brain region connection reported in the scientific literature does not necessarily reflect the physiological intensity of a connection; the former reflecting the interest for the region.

---

[1]Confidence (precision) has been evaluated for each extractor.

TABLE 2 | Statistics about corpus, extracted brain regions and connections (reproduced from Richardet et al., 2015).

| Corpus | Documents (words) | Brain regions mentions | | Connections mentions | |
|---|---|---|---|---|---|
| | | Lexical | Machine-learning | Lexical | Machine-learning |
| All PubMed abstracts | 13,293,649 ($2.1 \times 10^9$) | 1,705,549 | 1,992,747 | 41,965 | 188,994 |
| Full text neuroscience articles | 630,216 ($6.1 \times 10^9$) | 2,327,586 | 2,751,952 | 62,095 | 279,100 |

TABLE 3 | Brain regions for which connections have been found in the literature for the globus pallidus, internal segment and the subthalamic nucleus using text-mining models.

| Globus pallidus internus | | Subthalamic nucleus | |
|---|---|---|---|
| Region | Score | Region | Score |
| Caudoputamen | 143 | Globus pallidus, external segment | 105 |
| Globus pallidus, external segment | 117 | Caudoputamen | 74 |
| Pallidum | 23 | Cerebral cortex | 43 |
| Substantia nigra, reticular part | 21 | Pallidum | 34 |
| Subthalamic nucleus | 20 | Pedunculopontine nucleus | 16 |
| Lateral habenula | 12 | Thalamus | 16 |
| Thalamus | 10 | Globus pallidus, internal segment | 15 |
| internal capsule | 7 | Primary motor area | 11 |
| Cerebral cortex | 4 | Somatomotor areas | 9 |
| Hypothalamus | 3 | Substantia nigra, reticular part | 9 |
| Substantia nigra, compact part | 3 | Parafascicular nucleus | 7 |
| Pedunculopontine nucleus | 2 | Zona incerta | 5 |
| Cerebellar nuclei | 2 | Substantia nigra, compact part | 5 |
| Midbrain | 2 | Ventral tegmental area | 3 |
| Parafascicular nucleus | 2 | Midbrain | 2 |
| Lateral preoptic area | 2 | Lateral hypothalamic area | 2 |
| Cerebellum | 1 | Hypothalamus | 2 |
| Reticular nucleus of the thalamus | 1 | Brain stem | 2 |
| internal medullary lamina of the thalamus | 1 | Pons | 1 |
| Striatum-like amygdalar nuclei | 1 | internal medullary lamina of the thalamus | 1 |
| Zona incerta | 1 | Red nucleus | 1 |
| stria medullaris | 1 | striatonigral pathway | 1 |
| Fields of Forel | 1 | Isocortex | 1 |
| Magnocellular nucleus | 1 | Dentate nucleus | 1 |
| Central lateral nucleus of the thalamus | 1 | Substantia innominata | 1 |
| Claustrum | 1 | Bed nuclei of the stria terminalis | 1 |
| Substantia innominata | 1 | Islands of Calleja | 1 |
| Brain stem | 1 | Dorsal nucleus raphe | 1 |
| nigrostriatal tract | 1 | Cerebral nuclei | 1 |
| Interbrain | 1 | Olfactory tubercle | 1 |
| optic tract | 1 | Auditory areas | 1 |
| Ammon's horn | 1 | | |

All the results including suggested articles, nucleus and scores can be found in http://connectivity-brainer.rhcloud.com.

For the "Globus pallidus, internal segment," all LIT targets have been correctly suggested by the TM algorithm using ABA lexicon, except for one, ventral tegmental area, VTA. However, VTA is correctly proposed while searching using

ABA or BraiNER for "Pallidum" or "Pallidum, ventral region" instead of globus pallidus, internal segment. The result can be checked in http://connectivity-brainer.rhcloud.com/static/br/search.html.

TM proposes more targets for the GPi than the manual literature review, including connections with hypothalamus (3 publications), cerebellar nuclei (2), midbrain (2), parafascicular nucleus (2), and lateral preoptic area (2). The majority of the suggested targets includes or belongs to targets resulted from the manual literature review: midbrain includes SN; parafascicular nucleus relates to thalamus. However some of the targets proposed by TM were not found by LIT. Analyzing one such abstract suggested by TM, globus pallidus connection to the hypothalamus, the parafascicular nucleus and the lateral preoptic area are explicitly reported. TM found confirmatory sentences for the previously mentioned connections: ≪ *On the other hand, the dense substance P-positive wooly-fiber plexus filling the internal pallidal segment (entopeduncular nucleus) expands medialward into the lateral hypothalamic region.* ≫ or ≪ The **entopeduncular nucleus** invades the **hypothalamus** also with a loose plexus of enkephalin-positive wooly fibers ≫ (Haber and Nauta, 1983). For connections with the cerebellar nuclei, TM suggests papers that were not found by LIT, but these papers do not contain evidence of a connection. For illustration, we found three sentences that do not contain evidence of a connection with the cerebellar nuclei and all of them concern the cat. One example is ≪ Seventy *seven thalamic neurons in the VA-VL nuclear complex of the cat which projected to the anterior sigmoid gyrus (ASG) were studied extracellularly, and their responses to stimulation of both the cerebellar nuclei (CN) and the entopeduncular nucleus (ENT) were examined.* ≫ (Jinnai et al., 1987). This sentence is an example of a coordinating conjunction (e.g., ≪ Region A *and* Region B were examined. ≫). It was suggested by the simplest TM model that is not capable of filtering out coordinating conjunctions (even though they very rarely represent a connection).

For the STN, all the LIT targets have been found by TM, except for specific subdivisions of a given, such as ipsilateral ventral thalamic nuclei, ventral pallidum or the anterior cingulate. However, less specific regions (thalamus, pallidum) are correctly proposed. In addition, when using the machine learning named entity recognizer, the connection between STN and the ventral pallidum, anterior cingulate and ventral lateral thalamus are found as shown in: http://connectivity-brainer.rhcloud.com/static/br/region.html?db=20140522_brainer&br=1922.

For NAcc, **Table 4** (left) lists brain regions for which connections have been found in the literature based on the ABAlex named entity recognizer. Additionally, **Table 4** (right) also includes results from BraiNER (machine learning named entity recognizer). As discussed in Section Text-Mining, BraiNER is not constrained on a list of brain regions (like ABAlex) and is able to identify complex brain region names, even if they are not present in a lexicon. However, the regions returned by BraiNER have to be manually identified and curated as provided by the following link http://connectivity-brainer.rhcloud.com/static/br/region.html?br=912&db=20140522_brainer.

All the LIT targets, except the subgenual and pregenual cortex, have been found by the TM with the exact terminology. The two exceptions are explained by the fact that they are subdivisions of the anterior cingulate that figures as target.

Overall, TM has a precision of 36%, meaning that it proposed three times as many targets as could be identified with LIT. Such a low precision is acceptable for the task at hand, since the priority is to suggest all targets (high recall), even if that requires manual curation of search results (since precision is only 36%) The overall recall of TM against LIT in our study was 98%, meaning that over all the targets for the three selected seeds, only one target have been missed by TM (Frontal eye field for the STN) (**Table 5**).

### Species Differentiation
**Table 6** lists the number of publications found by text mining, ordered by species. Species were identified using Linnaeus, a machine-learning model to identify species in biomedical text and resolve it to the NCBI taxonomy (Gerner et al., 2010). One interesting observation is the difference between the number of studies on NAcc in rat and in primates, demonstrating the little available information on NAcc connectivity coming from studies in primates including human

### Probabilistic Tractography
The targets for NAcc found during LIT and TM were used to perform tractography.

We selected one subject to illustrate the results of the DTI in the current manuscript.

**Figure 1** shows the strength of connectivity of NAcc to its targets by depicting the number of voxels within the NAcc that has a probability superior to 1% to be connected to a specific target.

Cortical targets such as the anterior and subcalosal cingulate, medial and lateral orbitofrontal cortex, ventrolateral prefrontal cortex, insula, gyrus rectus, olfactory cortex all exhibited connection to NAcc. Conversely, hippocampus and amygdala exhibited a lower probability of connection to NAcc than expected. Hypothalamus and thalamus and basal ganglia including caudate, putamen and pallidum well as STN exhibited a strong probability of connection. In agreement with previous knowledge, midbrain dopaminergic structures, SN and VTA exhibited high probability of connections with NAcc.

**Figure 2** shows the probabilistic tractography output from FSL for the nucleus accumbens, based on the pattern of connectivity. We built up the NAcc connectivity maps with the **associative map** corresponding to the **NAcc putative core** (probabilistic connectivity to the prefrontal cortex, including frontal superior, frontal middle and frontal inferior-pars opercularis, -pars triangularis) and, the lateral orbitofrontal (l-OFC) cortex) and the **limbic map** corresponding to the **NAcc putative shell** (with distribution to medial-orbitofrontal cortex (m-OFC), anterior cingulate cortex, subcalosal area (Brodman area 25), amygdala, hippocampus, habenula, hypothalamus and ventral tegmental area (Baliki et al., 2013).

TABLE 4 | The 25 brain regions with highest scores for which connections have been found in the literature for the nucleus accumbens based on ABA and braiNER lexicons.

| Nucleus accumbens | | | |
|---|---|---|---|
| ABA | | Brainer | |
| Region | Score | Region | Score |
| Ventral tegmental area | 454 | ventral tegmental area | 238 |
| Caudoputamen | 412 | Striatum | 95 |
| Cerebral cortex | 295 | prefrontal cortex | 68 |
| Striatum-like amygdalar nuclei | 175 | Amygdala | 54 |
| Hippocampal region | 122 | medial prefrontal cortex | 52 |
| Ammon's horn | 93 | Hippocampus | 47 |
| Hippocampal formation | 70 | Hippocampal | 41 |
| Pallidum | 61 | basolateral amygdala | 40 |
| Midbrain | 53 | caudate-putamen | 39 |
| Subiculum | 38 | Cortical | 35 |
| Thalamus | 28 | Mesolimbic | 31 |
| Hypothalamus | 28 | hippocampal formation | 29 |
| Periaqueductal gray | 23 | ventral pallidum | 26 |
| Olfactory tubercle | 22 | ventral striatum | 20 |
| Basolateral amygdalar nucleus | 19 | caudate putamen | 16 |
| Fimbria | 18 | Thalamus | 14 |
| Nucleus raphe pontis | 18 | Neostriatum | 13 |
| Entorhinal area | 18 | Septum | 13 |
| Dorsal nucleus raphe | 13 | caudate nucleus | 13 |
| Globus pallidus, external segment | 12 | Mesencephalic | 13 |
| medial forebrain bundle | 11 | Amygdaloid | 12 |
| Paraventricular nucleus of the thalamus | 11 | Limbic | 12 |
| Lateral preoptic area | 9 | dorsal raphe nucleus | 11 |
| Nucleus of the solitary tract | 8 | paraventricular of the thalamus | 11 |
| stria terminalis | 8 | corpus striatum | 11 |

*The complete results can be found in http://connectivity-brainer.rhcloud.com.*

TABLE 5 | Overall performance of TM against LIT.

|  | Found by LIT | Proposed by TM | Missed by TM | Precision | Recall |
|---|---|---|---|---|---|
| GPi | 10 | 32 | 0 | 0.31 | 1.00 |
| STN | 23 | 31 | 1 | 0.76 | 0.96 |
| Nucleus Accumbens | 21 | 85 | 0 | 0.24 | 1.00 |
| Overall | 54 | 148 | 1 | 0.36 | 0.98 |

## Discussion

An exponentially growing amount of data is being produced and published in neuroscience, propelled by improvements in existing and new measurement recording technologies (Brown, 2007; Schierwagen, 2008). This staggering growth represents a major challenge to identify useful information and do not lack valuable information (Balan et al., 2014). Much legacy information about neural connections is inaccurate or is misleading because it is vastly oversimplified and must be evaluated critically since brain circuitry has been examined with a succession of increasingly reliable methods Already available BAMS (Bota et al., 2003) have been designed and implemented for storing and manipulating structural data about the nervous system in text- and table-based format allowing searching by region name, species and references (author, source, year) (Bota and Arbib, 2004).

In this article, we proposed to assess text-mining (TM) models to automatically suggest targets from the neuroscientific literature for tractography studies. Many publications deal with DTI limitations (Hilgetag et al., 2000; Lin et al., 2001; Mori and van Zijl, 2002; Parker et al., 2002; Ciccarelli et al., 2003a,b; Kinoshita et al., 2005; Johansen-Berg and Behrens, 2006; Behrens et al., 2007; Jbabdi and Johansen-Berg, 2011; Campbell and Pike, 2014; Thomas et al., 2014). One of them is related to DTI outputs that are not yet fully validated, inaccurate (Thomas et al., 2014) and difficult to quantify with a potential underestimate of the fiber tracts, as mentioned previously, when compared to electrophysiological tests (Lin et al., 2001; Parker et al., 2002; Ciccarelli et al., 2003a,b; Kinoshita et al., 2005). Other limitations of tractography are related to resolution, termination criteria, the effect of noise on the accuracy of the tracking and partial volume effects (Mori and van Zijl, 2002). The termination criteria correspond to the inability from tractography to determine the precise origin/termination of connections in the cortex (Jbabdi and Johansen-Berg, 2011) and to detect synapses. Accuracy quantification and error detection are also limitations of tractography, unable to provide any confidence scores on the output results even if efforts are being made to improve imaging techniques and algorithms (Hilgetag et al., 2000; Behrens et al., 2007). Tractography is unable to tell whether an axon is afferent or efferent (Jbabdi and Johansen-Berg, 2011). However, although current tractography methods have limitations, the ability to localize fiber bundles is of great help to understand connections and structural organization of the human brain. Anatomical knowledge can be used to impose constraint in the tract reconstruction, thereby effectively reducing the likelihood of the occurrence of erroneous results. Even if this approach is applied to anatomically well-documented tracts (Mori and van Zijl, 2002), it is essential to validate probabilistic results and in particular in DBS, to explore a specific seed by studying patterns of connectivity, sub-parcellation and confirmation of functional zones (Parker et al., 2002; Ciccarelli et al., 2003b; Kinoshita et al., 2005; Johansen-Berg and Behrens, 2006; Barkhoudarian et al., 2010; Lakhan and Callaway, 2010; Sedrak et al., 2010; Traynor et al., 2010; Coenen et al., 2011, 2012a; Pouratian et al., 2011; Taljan et al., 2011; Lambert et al., 2012; Chowdhury et al., 2013; Rozanski et al., 2014; Sweet et al., 2014a). Brain structures as nucleus accumbens, are less documented in human. We believe that TM approaches can help neuroscientist to use the provided information to identify targets for tractography and document them in human. Two well-established DBS targets for movement disorders have been studied (GPi and STN) and, NAcc, an exploratory DBS target for psychiatric disorders. The output of the TM method was compared with the output of a manual, systematic review of the literature and the output of the probabilistic tractography using NAcc as seed structure. The concordance with data from manual search is significant and robust. The overall performance of the TM algorithm against manual literature review (LIT) in our study was 98% recall, meaning that almost all regions found with LIT were also proposed by TM. In particular, when compared with the systematic search of the literature, for the "Globus pallidus, internal segment," all LIT targets but one (VTA) have been correctly suggested when using the restricted ABA lexicon. This missing target could be recovered when using the machine learning named entity recognizer (BraiNER). For the STN, all the targets identified by manual literature review have been found with TM, except for subsequent divisions of a given target, identified (again) when using BraiNER. For NAcc, all the targets, except for the subdivisions of the anterior cingulate cortex have been identified. Overall and as expected, TM returns and proposes more targets than manual literature review, but also provides indication for the plausibility of a given connection between two regions. As an example, the connection between GPi and the Caudoputamen has a score of 143, making the connection highly probable. In contrast, only one single article has been found for the connection between GPi and Ammon's horn (Hippocampus).

TABLE 6 | Number of publications and percentage for which connections have been found for the 3 nuclei by species using text-mining.

| | NAcc | | STN | | GPi | |
|---|---|---|---|---|---|---|
| Species | Number of publications | Percentage | Number of publications | Percentage | Number of publications | Percentage |
| Rattus | 1572 | 45.1 | 198 | 29.7 | 260 | 41.9 |
| Mus | 133 | 3.8 | 14 | 2.1 | 10 | 1.6 |
| Homo Sapiens | 83 | 2.4 | 34 | 5.1 | 13 | 2.1 |
| Simiiformes | 23 | 0.7 | 12 | 1.8 | 2 | 0.3 |
| Chordata | 72 | 2.1 | 12 | 1.8 | 15 | 2.4 |
| Felidae | 36 | 1.0 | 21 | 3.2 | 54 | 8.7 |
| Canis | 17 | 05 | 3 | 0.5 | 20 | 3.2 |
| No species found | 1550 | 44.5 | 372 | 55.9 | 247 | 39.8 |

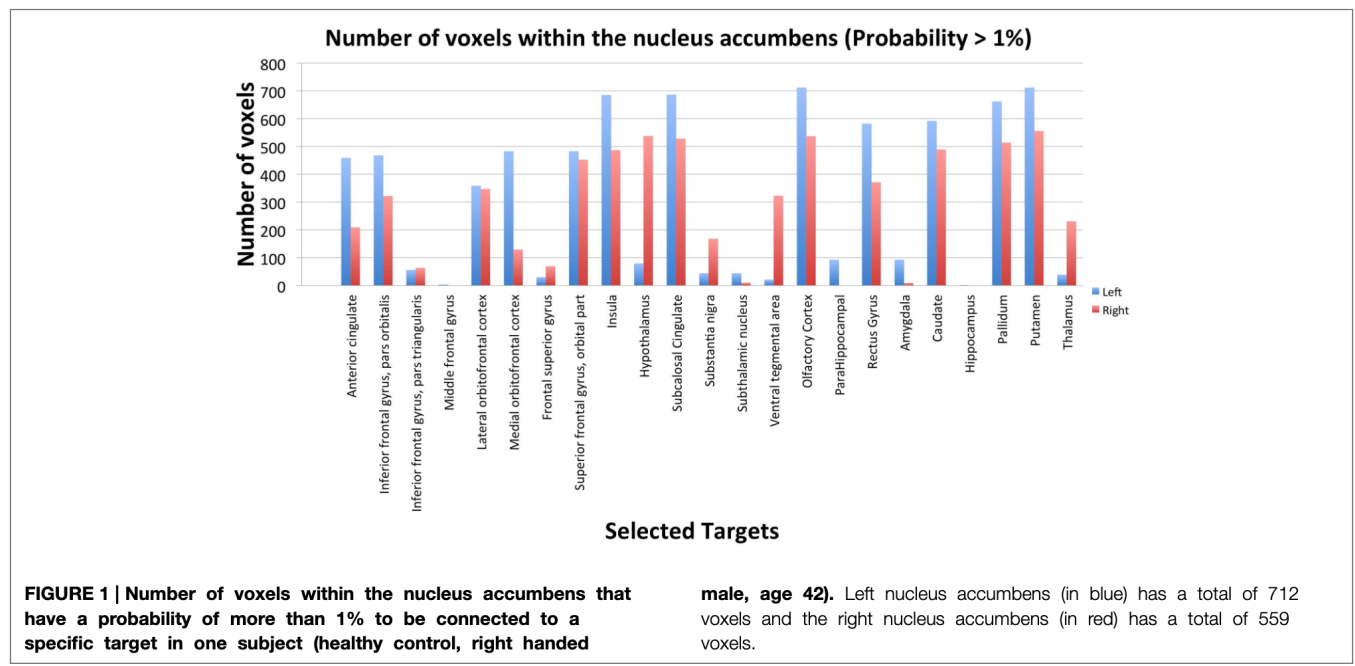

FIGURE 1 | Number of voxels within the nucleus accumbens that have a probability of more than 1% to be connected to a specific target in one subject (healthy control, right handed male, age 42). Left nucleus accumbens (in blue) has a total of 712 voxels and the right nucleus accumbens (in red) has a total of 559 voxels.

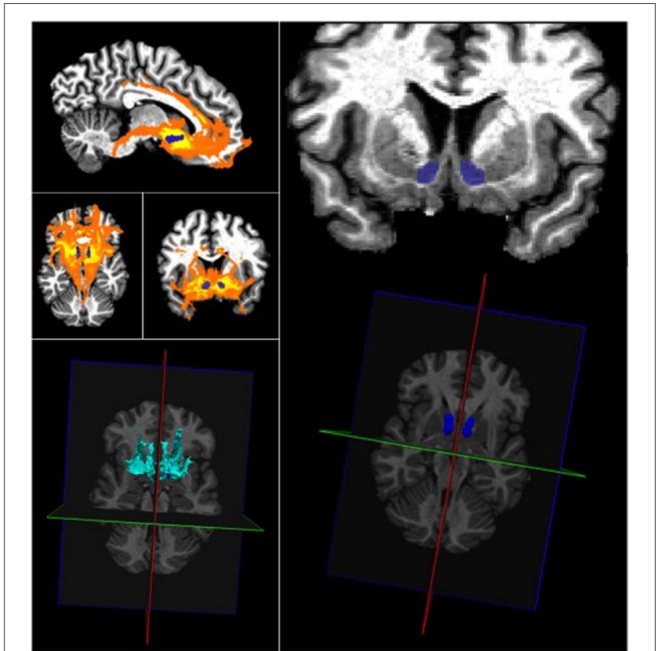

**FIGURE 2 | At the top left, the probabilistic tractography output of FSL with the left and right accumbens (in blue) on sagittal, axial and frontal slices (healthy control, right handed male, age 42).** Tracking the fibers passing through the nucleus accumbens with multi-fiber (3) tractography. A sagittal, axial and coronal maximum intensity projection is shown (yellow-orange). Bottom left: 3D view of the probabilistic tractography output. A 3D maximum intensity projection is shown in cyan with an axial MRI. On the right side, at the top, the identification of the left and right accumbens (blue) on coronal slice and at the bottom, the identification of the left and right accumbens in 3D (blue).

The key advantage of TM is the ability to screen millions of documents and billion of words in a matter of hours. This way, the *complete* available biomedical literature can be processed and analyzed. Another advantage is the possibility to search within results, and order them according to relevance. It is also possible to provide feedback to the models and subsequently retrain them with that additional data in order to improve results. However, TM has several shortcomings and manual post-processing of results is mandatory. For example, complex sentences are tedious to analyze and often yield incorrect or empty results. In fact, one has to keep in mind that the estimated precision of the proposed target regions by TM is 36%. TM is not yet able to extract the directionality of the connection, nor metadata like neurotransmitter type or if the connection is inhibitory or excitatory. Additionally, TM lacks the ability to clearly differentiate between facts and hypothesis and is not yet able to trace the source of a connectivity statement (e.g., when an articles cites another reference).

When compared to the TM models, the manual, systematic search of the literature has the major advantage to select and interpret data in the light of the known anatomy, resulting in a deep and thorough analysis of the available literature. Researchers are able to filter, synthetize and aggregate very disparate and complex information into a consistent knowledge base. They are capable of interpreting every connectivity statement, of replacing it in its specific context (including experimental setting, field of expertize of the authors), and therefore of judging the exact pertinence of a connectivity statement. This detailed manual analysis comes at the cost of scaling, meaning that only a fraction of the published data will be considered.

Obviously, both approaches have compelling advantages. However, we found that the winning strategy is to *combine* and leverage the strength of *both* approaches. Indeed: TM can be deployed as a first step to screen and aggregate the scientific literature, capable of ingesting millions of documents. Thereafter comes the time for a manual and meticulous analysis and verification of the suggested connectivity statements, with the possibility to drill down to the original source (published article). The manual effort can be directed on intelligent tasks like validating and searching proposed connectivity statement, instead of their painstakingly identification from within millions of publications. Using this dual strategy (TM prior to manual review), it took less than 2 h to have proposed a set of 25 potential targets for NAcc. In comparison, it took approximately a week for a user trained in neuroanatomy to conduct the isolated literature review of NAcc as presented in Section Manual Literature Review. Therefore, the connectivity database significantly accelerates the manual search of metascale brain region connectivity, by providing a centralized repository of connectivity data for neuroscientists. Another advantage of this dual approach is the possibility for neuroscientist to collectively curate a knowledge base and therefore improve it.

Regarding the distinction of connectivity statements from different species: as demonstrated by the review for the NAcc, the majority of the available data comes from rodent studies (Berendse et al., 1992a; Zahm, 1999; Van Kuyck et al., 2007). There is a striking need to disentangle human data from non-human primate data (Brauer et al., 2000). Frequently, information reported in humans is inferred from animal studies without further notice (Meredith et al., 1996). As provided by the results section, there is no sharp correspondence for the nomina between species for a given structure (e.g., globus pallidus, internal segment) rendering inferences from specie to another highly risky.

Furthermore, the pattern of connectivity for a given structure may differ between species (Ramnani et al., 2006; Bohland et al., 2009). Whether significant connections are reported between NAcc, hippocampus and amygdala through the available literature as identified via manual search and suggested by TM, the strength of connections between the aforementioned structures as output of the probabilistic tractography in healthy controls is not confirmatory of this result. A similar observation was reported for the subthalamic nucleus by Accolla et al. (2014). However, there are many examples of fiber pathways that are reported in dissection and tracer studies that are lacking in diffusion tensor tractography studies (Behrens et al., 2007), highlighting the importance of the selected tractography technique, its limitations and the potential role of the TM in validating connectivity information and support further investigations.

The design of an integrated platform where neuroscientist can access and curate proposed connectivity statements and share knowledge, using a standardized approach will provide significant new insights to neuroscience research. Early understanding on how to shape the TM can inform the design of future tools for neuroscience. The mining of large volumes of data and existing publications to identify patterns of and relationships between data from different levels of biological organization could help to predict parameters for experimental data to test and calibrate model implementations. Data curation and standardization is critically important to answer to brain modeling efforts as targeted by the Human Brain Project (Markram, 2012). One of the HBP objectives is to make it easier for neuroscientists to organize and access the massive volumes of heterogeneous data, knowledge and tools produced by the international neuroscience community. There is a need to bring together data from the literature, and from on-going research, and to provide a single source of annotated, high quality data.

Neuroscience is an incredibly diverse field with researcher coming from many disciplines. The cognitive psychologist might refer to Brodmann area 4, while the behavioral neuroscientist might refer to the primary motor cortex (Buitelaar et al., 2005). A researcher would not be disturbed by the different terminologies but a computer is. Furthermore, a researcher needs to have an overview of the existing difficulties posed by text before deciding on how to deal with. This is why curation process and standardization is crucial to fine-tune the TM outputs.

Which ontologies are used is also of major importance to ensure semantic heterogeneity when extracting information from various text sources. As we have seen, different instances of a region name can be used in publications which make the processing more complicated (Buitelaar et al., 2005; Ambert and Cohen, 2012). Several initiatives are trying to standardize neurosciences such as the International Neuroinformatics Coordinating Facility (INCF; http://www.incf.org/) with a global approach and more specifically Neuronames (Bowden and Martin, 1995; Bowden and Dubach, 2003) or the Neuroscience Information Framework (http://www.neuinfo.org) to fulfilling the need for standardized terminologies in neurosciences.

These techniques will provide predictions of fundamental importance for brain modeling in the operational phase of the project (Markram, 2012).

In the current study, we focused on the target identification using TM for tractography studies. TM improvements are also needed for the specificity of tractography applications, to visualize and explore projections extracted from the literature on a 3D atlas, to better evaluate topology, and speed up evaluation of results.

We believe that the TM approach could be useful for neuroscientists exploring specific DBS targets. DBS is one application but we also think that the text mining approach should be useful in helping the neuroscience community to facilitate global connectivity studies and in particular brain regions (Jbabdi and Johansen-Berg, 2011; Sporns, 2011). The applications of TM can be numerous in computational anatomy studies and in functional imaging in healthy and diseased brain. TM has also wide variety of applications in neuroscience (Tirupattur et al., 2011). The identification of biological entities such as protein and genes names as well as chemical compounds and drugs in free text, the association of gene clusters by microarray experiments with the biological context provided by the literature, automatic extraction of protein interactions and associations of proteins to functional concepts.

In conclusion, we demonstrate that connectivity for a structure of interest can be extracted from a very large amount of publications and abstracts. We believe this kind of approach will be useful in helping neuroscience community to facilitate connectivity studies of particular brain regions. The text mining tools used for the present study are indeed part of the HBP Neuroinformatics Platform and are freely available for the neuroscience community.

## Acknowledgments

We acknowledge our colleagues, Antoine Lutti, Sara Lorio, Anne Ruef and Lester Melie-Garcia for the validation and implementation of the structural imaging protocols at LREN. The research leading to these results has received funding from the European Union Seventh Framework Program (FP7/2007-2013) under grant agreement n° 604102 (Human Brain Project).

## Supplementary Material

The Supplementary Material for this article can be found online at: http://journal.frontiersin.org/article/10.3389/fnana.2015.00066/abstract

**Supplementary Figure 1 | Brain regions co-occurrences matrix displaying the top 20 regions for which the most connection mentions was found.** Matrix values represent the number of connectivity events, normalized by the confidence that each event has been extracted correctly (precision). All matrix values are linked to the corresponding detailed list of article sentences (see **Supplementary Figure 3**). The corresponding url for that figure is http://connectivity-brainer.rhcloud.com/static/br/matrix.html?db=20140226_aba&size=20.

**Supplementary Figure 2 | Listing of brain regions potentially connected to Nucleus accumbens, for which connectivity events have been found in the literature.** The score represents the number of connectivity events, normalized by the confidence that each event has been extracted correctly (precision). All regions are linked to the corresponding detailed list of article sentences (see **Supplementary Figure 3**). The corresponding url for that figure is http://connectivity-brainer.rhcloud.com/static/br/region.html?br=56&db=20140226_aba.

**Supplementary Figure 3 | Detailed list of sentences from neuroscientific articles, in this case between "Periaqueductal gray" and "Nucleus accumbens" (list truncated for readability).** Each sentence is linked to the original article on PubMed. Additionaly, the user has the ability to provide feedback: clicking on the red icon (thumbs down) will remove that sentence, and log it into the database. Similarly, clicking on the green icon (thumbs up) will confirm that sentence and log it in the database. The corresponding url for that figure is http://connectivity-brainer.rhcloud.com/static/br/details.html?br1=795&br2=56&db=20140226_aba/.

**Conflict of Interest Statement:** The authors declare that the research was conducted in the absence of any commercial or financial relationships that could be construed as a potential conflict of interest.